\documentclass[runningheads]{llncs}
\usepackage[T1]{fontenc}
\usepackage{graphicx}
\usepackage{amsmath}
\usepackage{amssymb}
\usepackage{array}
\usepackage[shortlabels]{enumitem}
\usepackage{hyperref}
\usepackage[nameinlink]{cleveref}
\usepackage{color}

\def\totp{\mathsf{TotP}}
\def\sp{\mathsf{\#P}}
\def\gapp{\mathsf{GapP}}
\def\gappl{\mathsf{GapP_{+}}}
\def\spanp{\mathsf{SpanP}}

\def\np{\mathsf{NP}}
\def\conp{\mathsf{coNP}}
\def\up{\mathsf{UP}}
\def\coup{\mathsf{coUP}}
\def\spp{\mathsf{SPP}}
\def\pp{\mathsf{PP}}
\def\cp{\mathsf{C_{=}P}}

\def\fp{\mathsf{FP}}
\def\fpl{\mathsf{FP_{+}}}
\def\p{\mathsf{P}}
\def\npsv{\mathsf{NPSV_{t}}}
\def\upsv{\mathsf{UPSV_{t}}}
\def\am{\mathsf{AM}}
\def\coam{\mathsf{coAM}}

\def\bpp{\mathsf{BPP}}

\newtheorem{Theorem}{Theorem}[section]
\newtheorem{Lemma}[Theorem]{Lemma}
\newtheorem{Proposition}[Theorem]{Proposition}
\newtheorem{Definition}[Theorem]{Definition}
\newtheorem{Corollary}[Theorem]{Corollary}

\begin{document}
	\title{Low Sets and Closure Properties of Counting Function Classes}

	\author{Yaroslav Ivanashev} %\orcidID{0009-0001-5059-1844}} %\and
		%Second Author\inst{2,3}\orcidID{1111-2222-3333-4444} \and
		%Third Author\inst{3}\orcidID{2222--3333-4444-5555}}
	%
	\authorrunning{Y. Ivanashev}
	
	\institute{HSE University, Faculty of Computer Science, Moscow, 101000, Russia
		\email{yivanashev@hse.ru}}
	%\institute{Princeton University, Princeton NJ 08544, USA \and
		%Springer Heidelberg, Tiergartenstr. 17, 69121 Heidelberg, Germany
	\maketitle              % typeset the header of the contribution
	\begin{abstract}
		A language $L$ is low for a relativizable complexity class $C$, if $C^{L} = C$. For the classes $\sp$, $\gapp$, and $\spanp$ the exact low classes of languages are known: $Low(\sp) = \up \cap \coup$, $Low(\gapp) = \spp$, and $Low(\spanp) = \np \cap \conp$. In this paper, we prove that $Low(\totp) = \p$, and give characterizations of low function classes for $\sp$, $\gapp$, $\totp$, and $\spanp$. In particular, we prove that $Low_{f}(\sp) = \upsv$ and $Low_{f}(\spanp)$\\ $= \npsv$. 
		We establish the inclusion relations between $\npsv$, $\upsv$, and the counting function classes by giving for each of these inclusions an equivalent inclusion between language classes. 
		%We establish the inclusion relations between $\npsv$, $\upsv$, and the counting classes by giving for each of the inclusions between these classes an equivalent inclusion between language classes. 
		We also prove that $\spanp \subseteq \gapp$ if and only if $\np \subseteq \spp$, and the inclusion $\gappl \subseteq \spanp$ implies $\mathsf{PH} = \mathsf{\Sigma^{P}_{2}}$. 
		For the class $\sp$ we prove that its closure under left composition with $\fpl$ is equivalent to $\sp = \upsv$, and for $\spanp$ this closure is equivalent to $\spanp = \npsv$. For the classes $\sp$, $\gapp$, $\totp$, and $\spanp$ we summarize the known results and show that each of these classes is closed under left composition with $\fpl$ if and only if it collapses to its low class of functions. We also prove that a NPTM with a $\sp$ oracle can always make at most one query to the oracle without changing the number of accepting paths.

		
	\end{abstract}
	\section{Introduction}
	
	The concept of lowness in complexity theory was introduced by Sch{\"o}ning to describe the languages in $\np$ that are low for the levels of polynomial-time hierarchy \cite{sch83}. A language $L$ is low for a relativizable complexity class $C$, if $C^{L} = C$. Informally, a language is low if it doesn't increase the computational power of a class when it is used as an oracle. For some complexity classes their low classes can be exactly  characterized. For example, $Low(\np) = \np \cap \conp$ \cite{sch83}, $Low(\am) = \am \cap \coam$ \cite{kob93}, and $Low(\bpp) = \bpp$ \cite{ko82}. For counting function classes $\sp$ \cite{val79}, $\gapp$ \cite{fen94}, and $\spanp$ \cite{kob89} the low classes are also known: $Low(\sp) = \up \cap \coup$ \cite{tor88,li93}, $Low(\gapp) = \spp$ \cite{fen94} and $Low(\spanp) = \np \cap \conp$ \cite{tor88}. In this paper, we give characterizations of a low language class for $\totp$ \cite{kia01} and low function classes for $\sp$, $\gapp$, $\gappl$, $\totp$, and $\spanp$. The results are summarized in \Cref{table}. In particular, we show that $Low_{f}(\sp) = \upsv$ and $Low_{f}(\spanp) = \npsv$. The classes $\npsv$ and $\upsv$ are the subclasses of $\mathsf{NPSV}$ and $\mathsf{UPSV}$ respectively, that contain only total functions \cite{boo84,boo85}. In \Cref{sec-rel}, we establish the relations between the classes $\npsv$, $\upsv$, and the counting function classes $\sp$, $\gappl$, $\totp$, and $\spanp$. We characterize all inclusions between these classes in terms of inclusions between language classes. In \Cref{sec-pre}, we establish the relations between classes $\gapp$ and  $\spanp$. We prove that $\spanp \subseteq \gapp$ if and only if $\np \subseteq \spp$, and the inclusion $\gappl \subseteq \spanp$ implies $\mathsf{PH} = \mathsf{\Sigma^{P}_{2}}$. The latter result implies that the inclusion $\sp \subseteq \gappl$ is proper unless $\mathsf{PH} = \mathsf{\Sigma^{P}_{2}}$.
	
	In \Cref{sec-fp}, we study the closure of counting function classes under left composition with $\fpl$. In \cite{ogi93}, it is shown that the closure of $\sp$ under composition with multivariate $\fpl$-functions is equivalent to $\pp = \up$, and for $\spanp$ this closure is equivalent to $\p^{\pp} = \np$ (in \cite{vol96}, this statement is extended to $\pp = \np$). We show that in these theorems univariate $\fpl$-functions can be considered, and the closure of $\sp$ is also equivalent to $\sp = \upsv$. For $\spanp$, we show that this closure is equivalent to $\spanp = \npsv$. Analogous theorems for $\gapp$ \cite{thi94,gup91,fen94} and $\totp$ \cite{iva25} are also known. In \cite{vol96}, Vollmer and Wagner proved that $\pp = \up$ (or, equivalently, $\cp = \up$ \cite{ogi93}) if and only if $\mathsf{FCH} = \sp$, which is equivalent to that $\sp$ is low for itself. \Cref{table} summarizes the results, which show that each of the classes $\sp$, $\gapp$, $\gappl$, $\totp$, and $\spanp$ is closed under left composition with $\fpl$ if and only if it coincides with its low class of functions.
	
	In \Cref{sec-sev}, we prove an auxiliary lemma, which shows that a composition of counting functions with multivariate $\fp$-functions is the same as a composition with univariate $\fp$-functions. We use this result in \Cref{sec-fp}. By a similar argument, we prove that a NPTM with a $\sp$ oracle can always make at most one query to the oracle preserving the same number of accepting paths. 
	
	\begin{table} %[t]
		\begin{center}
			\begin{tabular}{ | m{2.6cm} | m{2.6cm} | m{2.6cm} | m{3.8cm}| } 
				\hline
				Class & Low languages & Low functions & Closure under composition with $\fpl$ is equivalent to  \\ 
				\hline
				$\sp$ & $\up \cap \coup$ & $\upsv$ & $\sp = \upsv$ \newline $\pp = \up$ \\
				\hline
				$\gapp$ & $\spp$ & $\fp^{\spp}$ & $\gapp = \fp^{\spp}$ \newline $\pp = \spp$ \\
				\hline
				$\gappl$ & $\spp$ & $\fp^{\spp}$ & $\gappl = \fpl^{\spp}$ \newline $\pp = \spp$ \\
				\hline
				$\totp$ & $\p$ & $\fp$ & $\totp = \fpl$ \newline $\pp = \p$ \\
				\hline
				$\spanp$  & $\np \cap \conp$ & $\npsv$ & $\spanp = \npsv$ \newline $\pp = \np$  \\ 
				\hline
			\end{tabular}
		\end{center}
		\caption{Low classes of languages and functions, and conditions for the closure under left composition with $\fpl$ for the classes $\sp$, $\gapp$, $\gappl$, $\totp$, and $\spanp$.}
		\label{table}
	\end{table}
		
	\section{Preliminaries} \label{sec-pre}
	
	In this paper, we use standard notions of a deterministic and nondeterministic Turing machine. Below we give definitions of function and language classes that we will consider:
	
	\begin{Definition}
		\begin{enumerate}
			\item  \textnormal{\cite{val79}} $\sp = \{acc_{M} \ | \ M$ is a nondeterministic polynomial-time bounded Turing machine (NPTM)$\}$, where $acc_{M}(x)$ is the number of accepting paths of $M$ on input $x \in \Sigma^{*}$.		
			\item \textnormal{\cite{fen94}} $\gapp = \{gap_{M} \ | \ M$ is a NPTM$\}$, where $gap_{M}(x) = acc_{M}(x) - rej_{M}(x)$ and $rej_{M}(x)$ is the number of rejecting paths of $M$ on input $x \in \Sigma^{*}$.
			\item \textnormal{\cite{kia01}} $\totp = \{tot_{M} \ | \ M$ is a NPTM$\}$, where $tot_{M}(x) = Total_{M}(x) - 1$ and $Total_{M}(x)$ is the number of all paths of $M$ on input $x \in \Sigma^{*}$.
			\footnote{The "-1" in the definition of $\totp$ allows these functions to take a zero value. A predicate based definition of $\totp$ is given in \cite{iva25}.}		
			\item \textnormal{\cite{kob89}} $\spanp = \{span_{M} \ | \ M$ is a NPTM$\}$, where $span_{M}(x)$ is the number of distinct outputs on accepting paths of $M$ on input $x \in \Sigma^{*}$.		
			\item $\fp = \{f: \Sigma^{*} \rightarrow \mathbb{Z} \ | \ f$ is computable in polynomial time$\}$.		
			%\item $\fpl = \{f \ | \ f \in \fp$ and $f \geqslant 0\}$.
		\end{enumerate}
	\end{Definition}
	
	$\fpl$ and $\gappl$ are the subsets of $\fp$ and $\gapp$ respectively, that contain only non-negative valued functions. %of the form $\Sigma^{*} \rightarrow \mathbb{Z}_{\geqslant 0}$.
	
	\begin{Definition}
		\begin{enumerate}
			\item %\textnormal{\cite{coo71,lev73}}
			The class $\np$ consists of all languages $L$ for which there exists a function $f \in \sp$ such that for any $x \in \Sigma^{*}$: 
			\begin{align*}
				x \in L \Leftrightarrow f(x) > 0.
			\end{align*}
			\item %\textnormal{\cite{val76}} 
			The class $\up$ consists of all languages $L$ for which there exists a function $f \in \sp$ such that for any $x \in \Sigma^{*}$:
			\begin{align*}
				x \in L \Leftrightarrow f(x) = 1, \\
				x \notin L \Leftrightarrow f(x) = 0.
			\end{align*}
			\item %\textnormal{\cite{sim75,wag86}}
			The class $\cp$ consists of all languages $L$ for which there exist functions $f \in \sp$ and $g \in \fp$ such that for any $x \in \Sigma^{*}$:
			\begin{align*}
				x \in L \Leftrightarrow f(x) = g(x).
			\end{align*}
			\item %\textnormal{\cite{sim75,gil77}}
			The class $\pp$ consists of all languages $L$ for which there exist functions $f \in \sp$ and $g \in \fp$ such that for any $x \in \Sigma^{*}$:
			\begin{align*}
				x \in L \Leftrightarrow f(x) > g(x).
			\end{align*}
			\item %\textnormal{\cite{fen94}}
			The class $\spp$ consists of all languages $L$ for which there exists a function $f \in \gapp$ such that for any $x \in \Sigma^{*}$:
			\begin{align*}
				x \in L \Leftrightarrow f(x) = 1, \\
				x \notin L \Leftrightarrow f(x) = 0.
			\end{align*}
		\end{enumerate}
	\end{Definition}
	
	The following relations are well known or follow immediately from definitions:
	
	\begin{Proposition} \label[Proposition]{rel-bas}
		\begin{enumerate}
			\item $\fpl \subseteq \totp \subseteq \sp \subseteq \spanp$.
			\item \textnormal{\cite{fen94,bak24}} $\gapp = \sp - \sp = \sp - \fp = \totp - \totp = \totp - \fp$.
			\item $\p \subseteq \up \subseteq \np \subseteq \pp$.
			\item $\conp \subseteq \cp \subseteq \pp$.
			\item $\sp \subseteq \fp^{\pp}$.
		\end{enumerate}
	\end{Proposition}
	
	The next theorem gives characterizations of the inclusions between function classes that are not known to hold:
	
	\begin{Theorem}
		\begin{enumerate}
			%\item \textnormal{\cite{}}
			\item \textnormal{\cite{iva25}} $\sp = \totp$ if and only if $\p = \np$. 
			\item \textnormal{\cite{kob89}} $\spanp = \sp$ if and only if $\np = \up$. 
			\item $\spanp \subseteq \gapp$ if and only if $\np \subseteq \spp$. 
			\item If $\gappl \subseteq \spanp$, then $\mathsf{PH} = \mathsf{\Sigma^{P}_{2}}$.
			%\item If $\gappl \subseteq \sp$, then $\mathsf{PH} = \mathsf{\Sigma^{P}_{2}}$.
			\item The inclusion $\sp \subseteq \gappl$ is proper unless $\mathsf{PH} = \mathsf{\Sigma^{P}_{2}}$. 
		\end{enumerate}
	\end{Theorem}
	
	\begin{proof}
		\begin{enumerate} 
			\setcounter{enumi}{2}
			\item The proof from right to left is given in \cite[Corollary 3.5]{mah94}. The proof from left to right follows from \Cref{np-span}.
			\item For every language $L \in \cp$ there exists a function $f \in \gappl$ such that $L = \{x \ | \ f(x) = 0\}$ \cite{for97}. If $\gappl \subseteq \spanp$, then $L \in \conp$. From $\cp = \conp$ it follows that $\mathsf{PH} = \mathsf{\Sigma^{P}_{2}}$ \cite[Proposition 2.3.1]{ike22}: $\mathsf{PH} \subseteq \p^{\sp} \subseteq \np^{\sp} = \np^{\cp} = \np^{\conp} = \mathsf{\Sigma^{P}_{2}}$, where $\mathsf{PH} \subseteq \p^{\sp}$ is by Toda's theorem \cite{tod91}.
			\item Immediate from $(4)$. \qed
		\end{enumerate}
	\end{proof}
	
	\begin{Lemma} [\cite{tor88}] \label[Lemma]{np-span}
		A language $L$ is in the class $\np$ if and only if there exists a function $f \in \spanp$ such that for any $x \in \Sigma^{*}$:
		\begin{align*}
			x \in L \Leftrightarrow f(x) = 1, \\
			x \notin L \Leftrightarrow f(x) = 0.
		\end{align*}
	\end{Lemma}
	
	In \cite{tod92}, it is shown that $\sp^{\mathsf{PH}} \subseteq \fp^{\sp[1]}$. It follows that $\spanp \subseteq \sp^{\np} \subseteq \sp^{\mathsf{PH}} \subseteq \fp^{\sp[1]}$ \cite[Theorem 11]{kia01}. From \Cref{rel-bas} we have the following corollary: %\footnote{The class $\mathsf{SpanL}$ \cite{alv93}, which is a subset of $\totp$, is also known to be 1-Cook equivalent with $\sp$ \cite{alv93,hem95}. In \cite{hem95}, it is shown that $\gapp = \mathsf{SpanL} - \mathsf{FL}$. It follows that $\fp^{\mathsf{SpanL}[1]} = \fp^{\sp[1]} = \fp^{\spanp[1]}$.}
	
	\begin{Proposition} \label[Proposition]{tur-eq}
		$\fp^{\sp[1]} = \fp^{\gapp[1]} = \fp^{\totp[1]} = \fp^{\spanp[1]}$.
	\end{Proposition}
	
	\begin{Definition} [\cite{boo84,boo85}]
		\begin{enumerate}
			\item The class $\npsv$ consists of all functions $f$ such that there exists a NPTM that on any input $x \in \Sigma^{*}$ has at least one accepting path and outputs the value of $f(x)$ on every accepting path.
			\item The class $\upsv$ consists of all functions $f$ such that there exists a NPTM that on any input $x \in \Sigma^{*}$ has exactly one accepting path and outputs the value of $f(x)$ on the unique accepting path.
		\end{enumerate}
	\end{Definition}

	\begin{Proposition} \label[Proposition]{npsv-fp}
		\begin{enumerate}
			\item \textnormal{\cite{sel94}} $\npsv = \fp^{\np \cap \conp}$.
			\item \textnormal{\cite{hem06}} $\upsv = \fp^{\up \cap \coup}$.
		\end{enumerate}
	\end{Proposition}
	
	The proof of \Cref{npsv-fp} is given in \cite[Proposition 2.4]{boo85}. 
	
	In sections 4 and 5, we assume that the classes $\npsv$, $\upsv$ contain only non-negative integer valued functions in order to compare them with the integer-valued counting functions.
	
	\section{Low Languages and Low Functions}
	
	In this section, we characterize the classes of languages and functions that are low for counting function classes. 
	
	\begin{Definition}
		Let $C$ be a relativizable complexity class. 
		\begin{enumerate}
			\item A language $L$ is low for the class $C$, if $C^{L} = C$. $Low(C)$ is the class of all languages that are low for $C$.
			\item A function $g$ is low for the class $C$, if $C^{g} = C$. $Low_{f}(C)$ is the class of all functions that are low for $C$.
		\end{enumerate}
	\end{Definition}
	
	\begin{Theorem}
		\begin{enumerate} 
			\item \textnormal{\cite{tor88,li93}} $Low(\sp) = \up \cap \coup$.
			\item \textnormal{\cite{fen94}} $Low(\gapp) = \spp$.
			\item \textnormal{\cite{fen94}} $Low(\gappl) = \spp$.
			\item \textnormal{\cite{tor88}} $Low(\spanp) = \np \cap \conp$.
			\item $Low(\totp) = \p$.
		\end{enumerate}
	\end{Theorem}
	
	\begin{proof} 	
		\begin{enumerate} 
			\setcounter{enumi}{4}
			\item All languages in $\p$ are low for $\totp$. If a language $L$ is low for $\totp$, then there exists a function $f$ in $\totp^{L} = \totp$ such that for any $x \in \Sigma^{*}$:
			\begin{align*}
				x \in L \Leftrightarrow f(x) = 1, \\
				x \notin L \Leftrightarrow f(x) = 0.
			\end{align*}
			Then $L$ is in $\p$, because every polynomial bounded $\totp$-function is in $\fp$ \cite{bak24}. \qed
		\end{enumerate}
	\end{proof}
	
	In \Cref{low-functions} we assume that a NPTM can have as an oracle only a function such that the length of its output value is polynomially bounded in the length of the input.

	\begin{Corollary} \label[Corollary]{low-functions}
		\begin{enumerate} 
			\item $Low_{f}(\sp) = \upsv$.
			\item $Low_{f}(\gapp) = \fp^{\spp}$.
			\item $Low_{f}(\gappl) = \fp^{\spp}$.
			\item $Low_{f}(\totp) = \fp$.
			\item $Low_{f}(\spanp) = \npsv$.
		\end{enumerate}
	\end{Corollary}
	
	\begin{proof}
		We give a proof only for $\sp$. The proofs for $\gapp$, $\gappl$, $\totp$, and $\spanp$ are similar. $\upsv$ is low for $\sp$, because $\sp^{\upsv} = \sp^{\fp^{\up \cap \coup}} \subseteq \sp^{\up \cap \coup} = \sp$. For a function $f$, let %\footnote{The terms $graph(f)$ and $pregraph(f)$ are used in \cite{}.}
		$pregraph(f) = \{(x, y) \ | \ y$ is a prefix of $f(x)\}$ \cite{boo85}. If a function $f$ is low for $\sp$, then $\sp^{pregraph(f)} \subseteq \sp^{f} = \sp$, which implies that $pregraph(f) \in \up \cap \coup$. Therefore, $f \in \fp^{pregraph(f)} \subseteq \fp^{\up \cap \coup}$. \qed
	\end{proof}
	
	\section{Relations Between NPSVt, UPSVt, and Counting Classes} \label{sec-rel}
	
	In this section, we establish the inclusion relations between the classes $\npsv$, $\upsv$, and the counting function classes $\sp$, $\gappl$, $\totp$, and $\spanp$. The classes $\sp$, $\gappl$, $\spanp$ contain $\upsv$, and $\spanp$ contains $\npsv$ unconditionally. For the other inclusions we give characterizations in terms of inclusions between language classes.
	
	\begin{Proposition} \label[Proposition]{upsv-totp}
		$\upsv \subseteq \totp$ if and only if $\p = \up \cap \coup$.
	\end{Proposition}
	
	\begin{proof}
		Let $L$ be a language in $\up \cap \coup$. If $\upsv = \fp^{\up \cap \coup} \subseteq \totp$, then the function 
		\begin{align*}
			f(x) = \begin{cases}
				1, &\text{$x \in L$;}\\
				0, &\text{$x \notin L$.}
			\end{cases}
		\end{align*}
		is in $\totp$. Then $L$ is in $\p$, because every polynomial bounded $\totp$-function is in $\fp$ \cite{bak24}. 
		
		If $\p = \up \cap \coup$, then $\upsv = \fp^{\up \cap \coup} = \fp \subseteq \totp$. \qed
	\end{proof}
	
	\begin{Proposition}
		\begin{enumerate}
			\item $\npsv \subseteq \totp$ if and only if $\p = \np \cap \conp$.
			\item $\npsv \subseteq \sp$ if and only if $\up \cap \coup = \np \cap \conp$.
			\item $\npsv \subseteq \gappl$ if and only if $\np \cap \conp \subseteq \spp$.
		\end{enumerate}
	\end{Proposition}
	
	\begin{proof}
		The proofs from left to right are similar to \Cref{upsv-totp}. The proofs from right to left are immediate from $\fp \subseteq \totp$, $\fp^{\up \cap \coup} \subseteq \sp$, and $\fp^\spp \subseteq \gappl$. \qed
	\end{proof}
	
	\begin{Proposition} \label[Proposition]{in-upsv}
		The following statements are equivalent:
		\begin{enumerate}
			\item $\totp \subseteq \upsv$.
			\item $\sp \subseteq \upsv$.
			\item $\gappl \subseteq \upsv$.
			\item $\spanp \subseteq \upsv$.
			\item $\pp = \up$.
		\end{enumerate}
	\end{Proposition}
	
	\begin{proof}
		To prove that $(1 - 4)$ implies $(5)$, it suffices to show that $(1)$ implies $(5)$, because $\totp$ is a subset of $\sp$, $\gappl$, and $\spanp$. Let $L$ be a language in $\pp$. Then there exist functions $f \in \totp$ and $g \in \fp$ such that $L = \{x \ | \ f(x) > g(x)\}$ by \Cref{rel-bas}(2). If $\totp \subseteq \upsv$, then there exists a NPTM $M$ that on any input $x \in \Sigma^{*}$ has exactly one accepting path and outputs $f(x)$ on the unique accepting path. Then $L \in \up$, because there exists a NPTM $M'$ that simulates $M(x)$ and accepts on the unique accepting path of $M(x)$ if and only if $f(x) > g(x)$.
		
		$(5)$ implies $(1 - 4)$, because every class from $\sp$, $\totp$, $\gappl$, and $\spanp$ is a subset of $\fp^{\sp}$ by \Cref{tur-eq}. If $\pp = \up$, then $\fp^{\sp} = \fp^{\pp} = \fp^{\up \cap \coup} = \upsv$. \qed	
	\end{proof}
	
	\begin{Proposition}
		The following statements are equivalent:
		\begin{enumerate}
			\item $\totp \subseteq \npsv$.
			\item $\sp \subseteq \npsv$.
			\item $\gappl \subseteq \npsv$.
			\item $\spanp \subseteq \npsv$.
			\item $\pp = \np$.
		\end{enumerate}
	\end{Proposition}
	
	\begin{proof}
		The proof is similar to \Cref{in-upsv}. \qed
	\end{proof}
	
	\section{Closure Under Composition With FP} \label{sec-fp}
	
	In this section, we study the closure of counting function classes under left composition with the class $\fpl$. In \cite{ogi93}, it is shown that the closure of $\sp$ under composition with multivariate $\fpl$-functions is equivalent to $\pp = \up$, and for $\spanp$ this closure is equivalent to $\p^{\pp} = \np$ (in \cite{vol96}, this statement is extended to $\pp = \np$). We show that in these theorems univariate $\fpl$-functions can be considered, and also we add an equivalent statement $\sp = \upsv$ to \Cref{fp-sp}, and to \Cref{fp-span} we add a statement $\spanp = \npsv$. An implicit proof of equivalence between $\sp = \upsv$ and $\pp = \up$ can be found in \cite[Theorem 4.3(1)]{kos99} and \cite[Theorem 5.11(2)]{hem06}. We give the proofs for $\sp$ and $\spanp$, which can also be applied for $\gapp$, $\gappl$, and $\totp$.
	
	\begin{Theorem} \label{fp-sp}
		The following statements are equivalent:
		\begin{enumerate}
			\item $\fpl \circ \sp \subseteq \sp$.
			\item $\sp = \upsv$.
			\item $\pp = \up$.
		\end{enumerate}
	\end{Theorem}
	
	\begin{proof}	
		$(1 \Rightarrow 3)$: Let $L = \{x \ | \ f(x) > g(x)\}$, where $f \in \sp$, $g \in \fp$. The function 
		\begin{align*}
			h(x) = \begin{cases}
				1, &\text{$f(x) > g(x)$;}\\
				0, &\text{$\textnormal{otherwise}$.}
			\end{cases}
		\end{align*}
		is in $\fpl^{\sp[1]}$, which coincides with the class $\fp_{+} \circ \sp$ by \Cref{comp-sp}. By assumption, $h \in \sp$ and $L \in \up$. 
		
		$(3 \Rightarrow 2)$: If $\pp = \up$, then $\sp \subseteq \fp^{\pp} = \fp^{\up \cap \coup} = \upsv$.
		
		$(2 \Rightarrow 1)$: Follows from $\fpl \circ \upsv \subseteq \upsv$. \qed
	\end{proof}
	
	\begin{Theorem} \label{fp-span}
		The following statements are equivalent:
		\begin{enumerate}
			\item $\fpl \circ \spanp \subseteq \spanp$.
			\item $\spanp = \npsv$.
			\item $\pp = \np$.
		\end{enumerate}
	\end{Theorem}
	
	\begin{proof}
		$(1 \Rightarrow 3)$: Let $L = \{x \ | \ f(x) > g(x)\}$, where $f \in \sp$, $g \in \fp$. The function 
		\begin{align*}
			h(x) = \begin{cases}
				1, &\text{$f(x) > g(x)$;}\\
				0, &\text{$\textnormal{otherwise}$.}
			\end{cases}
		\end{align*}
		is in $\fpl^{\spanp[1]}$, which coincides with the class $\fp_{+} \circ \spanp$ by \Cref{comp-sev}. By assumption, $h \in \spanp$. By \Cref{np-span}, $L \in \np$. 
		
		$(3 \Rightarrow 2)$: If $\pp = \np$, then $\spanp \subseteq \fp^{\sp} = \fp^{\pp} = \fp^{\np \cap \conp} = \npsv$.
		
		$(2 \Rightarrow 1)$: Follows from $\fpl \circ \npsv \subseteq \npsv$. \qed
	\end{proof}
	
	The proof of equivalence between $\fpl \circ \gapp \subseteq \gapp$ and $\pp = \spp$ can be found in \cite{gup91,fen94}.
	
	\begin{Theorem} [\cite{thi94,gup91,fen94}]
		The following statements are equivalent:
		\begin{enumerate}
			\item $\fpl \circ \gapp \subseteq \gapp$.
			\item $\gapp = \fp^{\spp}$.
			\item $\pp = \spp$.
		\end{enumerate}
	\end{Theorem}
	
	The theorem for $\gappl$ can be proved similarly. In \cite{thi94}, it is shown that $\pp = \spp$ is equivalent to $\gappl = \fpl^{\spp}$.
	
	\begin{Theorem} [\cite{thi94,gup91,fen94}]
		The following statements are equivalent:
		\begin{enumerate}
			\item $\fpl \circ \gappl \subseteq \gappl$.
			\item $\gappl = \fpl^{\spp}$.
			\item $\pp = \spp$.
		\end{enumerate}
	\end{Theorem}
	
	The proof of the following theorem is given in \cite{iva25}.
	
	\begin{Theorem} [\cite{iva25}]
		The following statements are equivalent:
		\begin{enumerate}
			\item $\fpl \circ \totp \subseteq \totp$.
			\item $\totp = \fpl$.
			\item $\pp = \p$.
		\end{enumerate}
	\end{Theorem}
	
	\section{One Function Computes Several Functions} \label{sec-sev}
	
	In this section, we prove a technical lemma, which states that functions from $\sp$, $\gapp$, $\totp$, and $\spanp$ can output the values of several functions. The first application of this idea is \Cref{comp-sev}, which shows that a composition of $\fp$ with one function is the same as a composition with several functions. The second application is \Cref{sp-sp}, which states that a NPTM with a $\sp$ oracle can always make at most one query to the oracle without changing the number of accepting paths. A function $h$ is in the class $\fp \circ (\sp \times \ldots \times \sp)$ if there exist functions $f \in \fp$ and $g_{1}, ..., g_{k} \in \sp$ such that for any $x \in \Sigma^{*}$: $h(x) = f(g_{1}(x), \ldots, g_{k}(x))$. 
	
	\begin{Lemma} \label[Lemma]{comp-sp}
		$\fp \circ \sp = \fp^{\sp[1]} = \fp \circ (\sp \times \ldots \times \sp)$.
	\end{Lemma}
	
	\begin{proof}
		For every functions $f_{1}, \ldots, f_{k} \in \sp$ there exists a function $g \in \sp$ that on input $x \in \Sigma^{*}$ outputs a concatenation of the values of $f_{1}(x), \ldots, f_{k}(x)$. The function $g$ can be defined as $g(x) = f_{1}(x) \cdot 2^{(k - 1)p(|x|)} + f_{2}(x) \cdot 2^{(k - 2)p(|x|)} + \ldots + f_{k}(x)$, where $p$ is a polynomial such that for any $x \in \Sigma^{*}$: $|f_{1}(x)| \leqslant p(|x|), \ldots, |f_{k}(x)| \leqslant p(|x|)$. It follows that $\fp \circ \sp = \fp \circ (\sp \times \ldots \times \sp)$. $\fp^{\sp[1]}$ is a subset of $\fp \circ (\sp \times \fp)$, because for every function $f \in \fp^{\sp[1]}$ there exist functions $g, h \in \fp$ and $t \in \sp$ such that for any $x \in \Sigma^{*}$: $f(x) = g(t(h(x)), x)$. \qed
	\end{proof}
	
	For the classes $\gapp$, $\totp$, and $\spanp$ the same statement also holds, because they contain $\fpl$ and closed under addition, multiplication, and parsimonious reductions. From \Cref{tur-eq} we have the following corollary:
	
	\begin{Corollary} \label[Corollary]{comp-sev}
		$\fp \circ \sp = \fp \circ \gapp = \fp \circ \totp = \fp \circ \spanp = \fp^{\sp[1]} = \fp^{\gapp[1]} = \fp^{\totp[1]} = \fp^{\spanp[1]} = \fp \circ (\sp \times \ldots \times \sp) = \fp \circ (\gapp \times \ldots \times \gapp) = \fp \circ (\totp \times \ldots \times \totp) = \fp \circ (\spanp \times \ldots \times \spanp)$.
	\end{Corollary}
	
	In \cite[Lemma 4.3]{kob89}, it is proved that $\sp^{\np} = \sp^{\np[1]}$. We use a similar idea in \Cref{sp-sp}.
	
	\begin{Proposition} \label[Proposition]{sp-sp}
		\begin{enumerate}
			\item $\sp^{\sp} = \sp^{\sp[1]}$.
			\item $\gapp^{\sp} = \gapp^{\sp[1]}$.
		\end{enumerate}
	\end{Proposition}
	
	\begin{proof}
		\begin{enumerate}
			\item Let $M^{f}$ be a NPTM with an oracle $f \in \sp$.  There exists a NPTM $N^{g}$ with an oracle $g \in \sp$ such that for any $x \in \Sigma^{*}$: $acc_{M^{f}}(x) = acc_{N^{g}}(x)$ and on every computation path $N^{g}$ makes at most one query to the oracle. The machine $N^{g}$ simulates $M^{f}$ and guesses all oracle answers. Before it halts, $N^{g}$ verifies all oracle answers by a single query to the oracle, and rejects if at least one oracle answer was guessed incorrectly.	The oracle function $g$ is constructed so that on input $(x_{1}, \ldots, x_{k}) \in \Sigma^{*}$ the function $g$ outputs a concatenation of the values of $f(x_{1}), \ldots, f(x_{k})$. Let $S$ be a NPTM and $p$ be a polynomial such that for any $x \in \Sigma^{*}$: $acc_{S}(x) = f(x)$ and $|f(x)| \leqslant p(|x|)$. The function $g$ can be defined as $g(x) = acc_{T}(x)$, where $T$ is the NPTM that on input $(x_{1}, \ldots, x_{k}) \in \Sigma^{*}$: generates $k$ branches, simulates $S(x_{i})$ on the $i$-th branch for $1 \leqslant i \leqslant k$, and then on the accepting paths of $S(x_{i})$ generates $2^{(i - 1) \cdot p(\max(|x_{1}|, \ldots, |x_{k}|))}$ accepting paths. By construction, $g(x) = f(x_{k}) \cdot 2^{(k - 1)p(\max(|x_{1}|, \ldots, |x_{k}|))} + f(x_{k - 1}) \cdot 2^{(k - 2)p(\max(|x_{1}|, \ldots, |x_{k}|))} + \ldots + f(x_{1})$.
			\item The proof is similar to $(1)$. 
			\qed
		\end{enumerate}
	\end{proof}
	
	The same statement holds for languages that are defined using a single $\sp$ or $\gapp$ function:
	
	\begin{Corollary}
		\begin{enumerate}
			\item $\np^{\cp[1]} = \np^{\pp[1]} =  \np^{\pp} = \np^{\sp[1]} = \np^{\sp} = \np^{\gapp} = \np^{\totp} = \np^{\spanp}$.
			\item $\pp^{\cp[1]} = \pp^{\pp[1]} = \pp^{\pp} = \pp^{\sp[1]} = \pp^{\sp} = \pp^{\gapp} = \pp^{\totp} = \pp^{\spanp}$.
		\end{enumerate}
	\end{Corollary}

	%\subsubsection{\ackname}
	
	%The work is supported by the HSE Basic Research Program.
	
	\bibliographystyle{splncs04}
	\bibliography{low-sets-bibliography}

\end{document}